\documentclass[twocolumn,floatfix,amsmath,amssymb,preprintnumbers,superscriptaddress]{revtex4-2}

\usepackage{graphicx}% Include figure files
\usepackage{dcolumn}% Align table columns on decimal point
\usepackage{bm}% bold math
\usepackage{amsmath}
\usepackage{xcolor}
\usepackage{soul}
\usepackage{graphicx}% Include figure files
\usepackage{dcolumn}% Align table columns on decimal point
\usepackage{bm}% bold math
\def\lesssim{\ \raise.3ex\hbox{$<$}\kern-0.8em\lower.7ex\hbox{$\sim$}\ }
\def\gesim{\ \raise.3ex\hbox{$>$}\kern-0.8em\lower.7ex\hbox{$\sim$}\ }

\begin{document}
\preprint{NITEP 142}

\title{Resonance-to-bound transition of $^5$He in neutron matter and its analogy with heteronuclear Feshbach molecule}
\author{Hiroyuki Tajima}
\email{hiroyuki.tajima@tnp.phys.s.u-tokyo.ac.jp}
\affiliation{Department of Physics, Graduate School of Science, The University of Tokyo, Tokyo 113-0033, Japan}

\author{Hajime Moriya}
\email{moriya@nucl.sci.hokudai.ac.jp}
\affiliation{Department of Physics, Hokkaido University, Sapporo 060-0810,
  Japan}
\author{Wataru Horiuchi}
\email{whoriuchi@omu.ac.jp}
\affiliation{Department of Physics, Osaka Metropolitan University, Osaka 558-8585, Japan}
\affiliation{Nambu Yoichiro Institute of Theoretical and Experimental Physics (NITEP), Osaka Metropolitan University, Osaka 558-8585, Japan}
\affiliation{Department of Physics, Hokkaido University, Sapporo 060-0810,
  Japan}
\author{Kei Iida}
\email{iida@kochi-u.ac.jp}
\affiliation{Department of Mathematics and Physics, Kochi University, Kochi 780-8520, Japan}
\author{Eiji Nakano}
\email{e.nakano@kochi-u.ac.jp}
\affiliation{Department of Mathematics and Physics, Kochi University, Kochi 780-8520, Japan}

\date{\today}
\begin{abstract}
We theoretically investigate the fate of a neutron-alpha $p$-wave resonance in dilute neutron matter, which 
may be encountered in
neutron stars and supernova explosions.
While $^5$He is known as a resonant state that decays to a neutron and an alpha particle in vacuum,
this unstable state turns into a stable bound state in the neutron Fermi sea 
because the decay process is forbidden by the Pauli-blocking effect of neutrons.
Such a resonance-to-bound transition assisted by the Pauli-blocking effect can be 
realized
in cold atomic experiments for a quantum mixture near the heteronuclear Feshbach resonance.
\end{abstract}

%\begin{document}
\maketitle

%\tableofcontents
\section{Introduction}
\label{sec:1}
Nuclear clusters in astrophysical environments have gathered tremendous attention thanks to 
recent progress in multi-messenger astronomy.
Light cluster states such as a deuteron, a triton, and an alpha particle 
are expected to appear in supernova explosions as well as in neutron-star mergers~\cite{RevModPhys.89.015007}.
Moreover, multi-alpha clusters (e.g., the Hoyle state consisting basically 
of three alpha particles~\cite{hoyle1954resonances}), if appearing as bound states in such environments, would
play an important role in enhancing the nuclear reaction rate.
Recently, the existence of alpha clusters has also been confirmed in heavy nuclei~\cite{doi:10.1126/science.abe4688},
indicating the importance of the cluster states in medium.
Note that single- and multi-cluster states in medium
can be dramatically different from those in vacuum.
To connect such cluster states in vacuum, in nuclei, and in
astrophysical environments, therefore,
we need to consider 
a complicated many-body problem
involving inhomogeneities of the medium~\cite{HOROWITZ200655,PhysRevC.77.055804,PhysRevC.81.015803}.

A concept of polarons, which was originally proposed in the context of condensed matter physics~\cite{landau1948effective}
to tackle quasiparticle properties in many-body backgrounds, has been examined extensively and applied to cold atomic systems (for review, see, e.g., Ref.~\cite{massignan2014polarons}).
In particular, the controllable interaction associated with the Feshbach resonance~\cite{RevModPhys.82.1225} and the variety of atomic species with different quantum statistics
enable us to investigate many-body states of matter systematically as a quantum simulator~\cite{RevModPhys.80.885,RevModPhys.80.1215}.
In this regard, a quasiparticle impurity state immersed in a Fermi sea (Bose-Einstein condensate) called 
%as 
a Fermi (Bose) polaron
is of particular interest in cold atoms.
Moreover, utilizing the notion of such atomic polarons,
the present authors have elucidated the fate of alpha clusters, that is, $^8$Be and the Hoyle state, in dilute neutron matter~\cite{PhysRevC.104.065801}.
While $^8$Be and the Hoyle state are unstable two- and three-body alpha cluster states in vacuum,
it was found that they become bound states due to the polaronic properties of alpha particles (i.e., increased effective mass and mediated interpolaron interaction associated with $s$-wave neutron-alpha interaction)~\cite{PhysRevC.102.055802,atoms9010018}. 
On the other hand, a narrow $p$-wave resonance for neutron-alpha scattering 
is known to occur around $0.94$ {\rm MeV}~\cite{10.1143/PTP.61.1327},
which should be important at sufficiently high neutron densities. 
In fact, polarons and molecules predicted near the $p$-wave Feshbach resonance~\cite{PhysRevLett.109.075302,PhysRevA.100.062712} are analogous to alpha polarons and the emergence of 
$^5$He resonance and bound states in dilute neutron matter
in the sense that a $p$-wave quasiparticle state with finite lifetime turns into a bound state as coupling with medium changes effectively.
Also, in addition to the $s$-wave interaction examined in Ref.~\cite{PhysRevC.104.065801}, this $p$-wave interaction could have an increasing influence on the formation of multi-alpha clusters (e.g., $^9$Be~\cite{von1996two}).

\begin{figure}[t]
    \centering
    \includegraphics[width= \hsize]{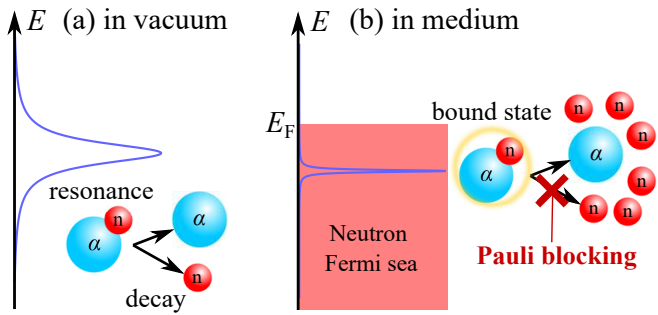}
    \caption{Schematics of (a) the $^5$He resonance decaying to a neutron and an alpha particle in vacuum and (b) the in-medium bound $^5$He in a neutron Fermi sea, where the decay process is forbidden by the Pauli blocking effect of neutrons.
    Such a sharp peak is in contrast with the broad peak found in the resonance.
    }
    \label{fig:1}
\end{figure}

In this paper, we focus on the neutron-alpha $p$-wave interaction that was not considered in the previous work~\cite{PhysRevC.102.055802,PhysRevC.104.065801}.
This interaction is associated with the $^5$He resonant state decaying to a neutron and an alpha particle.
While the $s$-wave neutron-alpha interaction is a leading contribution in the low-density limit of neutrons,
the $p$-wave resonance would not be negligible when the 
neutron Fermi energy 
exceeds the $^5$He resonance energy.
As shown in Fig.~\ref{fig:1},
we show that the $^5$He resonant state turns into a bound state because the decay process is forbidden by the Pauli-blocking effect of neutrons.
The situation is analogous to the
competition between molecules and polaron states
in the population-imbalanced mixture with the narrow Feshbach resonance.
Indeed, 
$p$-wave Feshbach molecules have been already observed experimentally~\cite{PhysRevLett.98.200403}, 
while $p$-wave polarons have been studied theoretically~\cite{PhysRevLett.109.075302,PhysRevA.100.062712}.
Moreover, the polaron-molecule transition for a narrow $s$-wave resonance has also been investigated in the literature~\cite{PhysRevA.85.053612,Massignan_2012}.
In this work, 
by describing
the in-medium $^5$He state 
within a
two-channel model, which is well established for the study of many-body atomic systems near the Feshbach resonance,
we draw an analogy 
of the resonance-to-bound transition of $^5$He in neutron matter
with heteronuclear Feshbach molecules coexisting with a Fermi sea in a Bose-Fermi mixture, where the transition from polaronic condensates to heteronuclear Feshbach molecules were recently observed~\cite{duda2021transition,duda2022long}.

This paper is organized as follows.
In Sec.~\ref{sec:2}, we introduce the two-channel Hamiltonian for the $^5$He resonance.
In Sec.~\ref{sec:3}, we examine the $p$-wave neutron-alpha scattering amplitude in our two-channel model.
In Sec.~\ref{sec:4}, we discuss the in-medium $^5$He state and its resonance-to-bound transition in neutron matter.
We summarize this paper in Sec.~\ref{sec:5}.

\par
\section{Hamiltonian}
\label{sec:2}
We consider a system of alpha particles, neutrons, and $^5$He nuclei by analogy with the $p$-wave Feshbach resonance in cold atomic systems.
We employ the two-channel Hamiltonian given by
\begin{align}
    H=H_{0,\nu}+H_{0,\alpha}+H_{0,\Phi}+V,
\end{align}
where
\begin{align}
    H_{0,\nu}
    &=\sum_{\bm{k}}\sum_{s_z=\pm1/2}\xi_{\bm{k},\nu}\nu_{\bm{k},s_z}^\dag \nu_{\bm{k},s_z},
\end{align}
\begin{align}
    H_{0,\alpha}
    &=\sum_{\bm{q}}\xi_{\bm{q},\alpha}\alpha_{\bm{q}}^\dag \alpha_{\bm{q}},
\end{align}
\begin{align}
    H_{0,\Phi}
    &=\sum_{\bm{P},J,J_z}\left(\xi_{\bm{P},\Phi,J}+E_{\Phi,J}\right)
    \Phi_{\bm{P},J,J_z}^\dag \Phi_{\bm{P},J,J_z}
\end{align}
are the kinetic terms of a spin-1/2 neutron, an alpha particles, and
a closed-channel neutron-alpha state (corresponding to the bare $^5$He state), respectively.
$\nu_{\bm{k},s_z}$ and $\alpha_{\bm{q}}$ denote the annihilation operators of a neutron and an alpha particle.
The annihilation operator $\Phi_{\bm{P},J,J_z}$ of the bare $^5$He state involves the total spin $J=3/2,1/2$ consisting of the neutron spin $s=1/2$ and the neutron-alpha relative angular momentum $\ell=1$, and its $z$-component $J_z$.
%$\sigma=\pm1/2$ is the $z$-component of $s$.
We have defined the kinetic energies (minus the chemical potentials) 
of neutrons, alpha particles, and closed-channel neutron-alpha states as
$\xi_{\bm{k},\nu}=\frac{k^2}{2M_{\nu}}-\mu_\nu$,
$\xi_{\bm{q},\alpha}=\frac{q^2}{2M_{\alpha}}-\mu_\alpha$,
and
$\xi_{\bm{P},\Phi,J}=\frac{P^2}{2M_\Phi}-\mu_{\Phi,J}$, 
respectively.
The energy level of the bare $^5$He state amounts to $E_{\Phi,J}$.
For simplicity, we consider the spin-degenerate system.
Moreover, we may use relations among masses given by $M_\Phi=M_{\nu}+M_{\alpha}$ and $M_\alpha=4M_{\nu}$.

The interaction term $V$ is given by
\begin{widetext}
\begin{align}
    V=\sum_{J=1/2,3/2}
%    \sum_{J_z=-J}^{j_z=+J}
    \sum_{J_z,s_z,m}
    \sum_{\bm{P},\bm{k}}
    \left(k{g}_{k,J}\sqrt{\frac{4\pi}{3}}Y_{m}^{\ell=1}(\hat{\bm{k}})
    \langle 1,m;1/2,s_z |J,J_z\rangle
    \Phi_{\bm{P},J,J_z}^\dag
    \nu_{\bm{k}+\bm{P}/2,s_z}\alpha_{-\bm{k}+\bm{P}/2}+{\rm h.c.}\right),
\end{align}
\end{widetext}
where $\langle 1,m;1/2,s_z |J,J_z\rangle$ is the Clebsch-Gordan coefficient. 
The $p$-wave properties manifest themselves in the Feshbach-like coupling $g_{k,J}$ with the $\ell=1$ component of the spherical harmonics $Y_{m}^{\ell=1}(\hat{\bm{k}})$, where $\bm{k}$ is the relative momentum between a neutron and an alpha particle.

%\begin{align}
%    H&=\sum_{\bm{k},\sigma}\xi_{\bm{k},\nu}\nu_{\bm{k},\sigma}^\dag \nu_{\bm{k},\sigma}
%    +\sum_{\bm{q}}\xi_{\bm{q},\alpha}\alpha_{\bm{q}}^\dag \alpha_{\bm{q}}\cr
%    &+\sum_{\bm{P},\sigma}\left(\xi_{\bm{P},\Phi}+E_\Phi\right)
%    \Phi_{\bm{P},\sigma}^\dag \Phi_{\bm{P},\sigma}\cr
%    &+\sum_{\bm{P},\bm{k},\sigma}g_{k}\bm{k}\left(
%    \nu_{\bm{k}+\bm{P}/2,\sigma}^\dag \alpha_{-\bm{k}+\bm{P}/2}^\dag 
%    \Phi_{\bm{P},\sigma}
%    +{\rm h.c.}
    %+
    %\Phi_{\bm{P},\sigma}^\dag \alpha_{-\bm{k}+\bm{P}/2} \nu_{\bm{k}+\bm{P}/2,\sigma}
%    \right),
%\end{align}
%where 
%$\nu_{\bm{k},\sigma}$, $\alpha_{\bm{q}}$, and $\Phi_{\bm{P},\sigma}$ denote the annihilation operators of a neutron, an alpha particle, and a closed-channel neutron-alpha state (corresponding to the bare $^5$He state), respectively.
%$g_k$ is the Feshbach-type $p$-wave coupling for the neutron-alpha scattering.

Hereafter we shall set $\mu_\alpha=0$ because alpha particles are supposed to be present as impurities.   For simplicity, we ignore the presence of electrons and keep the total number of neutrons and of protons unchanged.
The chemical potentials of neutrons, alpha particles, and $^5$He nuclei
have a relation given by $\mu_{\Phi,3/2}=\mu_{\Phi,1/2}\equiv\mu_{\Phi}$ and
$\mu_\nu+\mu_\alpha=\mu_\Phi$ because of the number conservation under thermodynamic equilibrium.

\section{$p$-wave neutron-alpha scattering amplitude}
\label{sec:3}
We proceed to examine the scattering properties in vacuum, 
i.e., at $\mu_\nu=\mu_{\alpha}=0$.
The scattering amplitude of the $\ell$-th partial wave with $J$ reads~\cite{PhysRevLett.94.090402,GURARIE20072}
\begin{align}
\label{eq:t-matrix}
     f_{\ell,J}(k)=-\frac{M_{\rm r}}{2\pi}T_{\ell,J}\left(k,k;\Omega=\frac{k^2}{2M_{\rm r}}\right),
\end{align}
where $M_{\rm r}=(1/M_\nu+1/M_\alpha)^{-1}$ is the reduced mass (note that $\Omega$ denotes the two-body energy).
$T_{\ell,J}(k,k';\Omega)$ is the $\ell$-th component of the neutron-alpha $T$-matrix at the zero center-of-mass frame.
$ f_{\ell=1,J}(k)$ is associated with the $p$-wave scattering phase shift $\delta_{\ell=1,J}(k)$ as
\begin{align}
\label{eq:phase-shift}
    f_{\ell=1,J}(k)=\frac{k^2}{k^3\cot\delta_{\ell=1,J}(k)-ik^3}.
\end{align}
The effective-range expansion of $\delta_{\ell=1,J}(k)$ reads
\begin{align}
\label{eq:effective-range}
    k^3\cot\delta_{\ell=1,J}(k)=-\frac{1}{a_J}+\frac{1}{2}r_Jk^2+O(k^3),
\end{align}
where $a_J$ and $r_J$ are the scattering volume and the effective range in the state with $J$, respectively.

To calculate the $p$-wave scattering amplitude in our model,
first we consider the bare closed-channel propagator $D_{J,J_z}^0(\bm{P},\Omega)$ with the momentum $\bm{P}$ and the energy $\Omega$ as
\begin{align}
    D_{J}^0(\bm{P},\Omega)=\frac{1}{\Omega_+-\xi_{\bm{P},\Phi,J}-E_{\Phi,J}},
\end{align}
where $\Omega_+=\Omega+i\delta$ with infinitesimally small value The renormalized propagator [physically corresponding to the ground ($J=3/2$) and first-excited ($J=1/2$) states of $^5$He] is given by
\begin{align}
    D_{J,J_z}(\bm{P},\Omega)
    &=\frac{1}{\left[D_{J}^0(\bm{P},\Omega)\right]^{-1}-\Pi_{J,J_z}(\bm{P},\Omega)},
\end{align}
where the one-loop self-energy $\Pi_{J,J_z}(\bm{P},\Omega)$ describes the decay process to a neutron and an alpha particle~\cite{Hammer_2017}.
At zero center-of-mass frame ($\bm{P}=\bm{0}$),
the self-energy is independent of $J_z$ as $\Pi_{J,J_z}(\bm{P}=\bm{0},\Omega)\equiv \Pi_{0,J}(\Omega)$
because of the spherical symmetry
in the momentum space given by
$\sum_{\bm{q}}q_j^2F(q)=\frac{1}{3}\sum_{\bm{q}}q^2F(q)$ for an arbitrary function $F(q)$ ($j=x,y,z$).
%Using the rotational symmetry and $q^2=q_x^2+q_y^2+q_z^2$, we rewrite it as
%$\Pi_{ij}(\bm{P},\Omega)=\delta_{ij}\Pi_{0}(\bm{P},\Omega)$, where
$\Pi_{0,J}(\Omega)$ reads
\begin{align}
\label{eq:piv}
    \Pi_{0,J}(\Omega)=
    \frac{1}{3}
    \sum_{\bm{q}}\frac{q^2g_{q,J}^2}{
    \Omega_+-\xi_{\bm{q},\nu}-\xi_{-\bm{q},\alpha}}.
\end{align}
Using $D_{J,J_z}(\bm{P}=\bm{0},\Omega)$ with $\Pi_{0,J}(\Omega)$,
we obtain the $p$-wave component of the $T$-matrix as~\cite{GURARIE20072,PhysRevLett.94.090402}
\begin{align}
\label{eq:tmatrix-propagator}
    T_{\ell=1,J}(k,k;\Omega)=\frac{1}{3}\frac{k^2g_{k,J}^2}{\Omega_+-E_{\Phi,J}-\Pi_{0,J}(\Omega)}.
\end{align}
%where we defined
%\begin{align}
%    T(k,k';\Omega)=\sum_{\ell=0}^{\ell=\infty}\sum_{m=-\ell}^{m=\ell}
%    T_{\ell,J}(k,k';\Omega)
%    Y_{m}^{\ell}(\hat{\bm{k}})Y_{m}^{\ell}(\hat{\bm{k}'}).
%\end{align}
%$p$-wave 
%$T$-matrix at the center-of-mass frame as
%\begin{align}
%\label{eq:tmatrix-propagator}
%    T(k,k',\Omega)&=\sum_{j=x,y,z}k_j k'_{j}
%    g_kg_{k'}D_{jj}(\bm{0},\Omega),
%\end{align}
%which is in turn related to the $p$-wave component $T_{\ell=1}(k)$ as~\cite{GURARIE20072}
%\begin{align}
%\label{eq:tmatrix-def}
%    T(k,k',\Omega)&=3T_{\ell=1}(k)
%    {\hat k}\cdot{\hat k'}.
%\end{align}
%In Eq.~(\ref{eq:tmatrix-def}), we have defined ${\hat k}=\bm{k}/k$ and ${\hat k'}=\bm{k}'/k'$.

Practically, we introduce the momentum-dependent coupling 
\begin{align}
\label{eq:gamma}
    g_{k,J}=g_J\gamma_{k,J}\equiv\frac{g_J}{1+(k/\Lambda_J)^2},
\end{align}
where $\Lambda_J$ is the cutoff scale and $g_{k,J}$ is assumed to be a real value.
While the step-like form factor $g_{k,J}=g_J\theta(\Lambda_J-k)$ is used in Ref.~\cite{PhysRevLett.109.075302},
a smoothly decreasing function $\gamma_{k,J}$ is required to reproduce the neutron-alpha scattering properties.
In this case, we can analytically obtain $\Pi_{0,J}(\Omega)$ as
\begin{align}
\label{eq:pi}
    \Pi_{0,J}(\Omega)
    &=-
    \frac{M_{\rm r}g_J^2\Lambda_J^4[\Lambda_J^3+6\Lambda_J M_{\rm r}\Omega+2i(2M_{\rm r}\Omega)^{\frac{3}{2}}]}{12\pi(2M_{\rm r}\Omega+\Lambda_J^2)^2}.
\end{align}
Combining Eqs.~(\ref{eq:t-matrix})-(\ref{eq:effective-range}) and Eqs.~(\ref{eq:tmatrix-propagator})-(\ref{eq:pi}), 
we obtain the relation between the low-energy constants (i.e., $a_J$ and $r_J$) and the model parameters (i.e., $g_J$, $E_{\Phi,J}$, and $\Lambda_J$) as
\begin{widetext}
\begin{align}
\label{eq:phase-shift_expanded}
&-\frac{1}{a_J}+\frac{1}{2}r_Jk^2-ik^3+O(k^4) 
    =-\frac{6\pi}{M_{\rm r}g_J^2}\left[1+\left(\frac{k}{\Lambda_J}\right)^2\right]^2
    %\cr
    %&\times
    \left[\frac{k^2}{2M_{\rm r}}-E_{\Phi,J}+
    \frac{M_{\rm r}g_J^2\Lambda_J^4(\Lambda_J^3+3\Lambda_J k^2+2ik^3)}{12\pi(k^2+\Lambda_J^2)^2}
    \right].
\end{align}
\end{widetext}
Comparing the coefficients of $k^0$ and $k^2$ between 
the left and right sides of Eq.~(\ref{eq:phase-shift_expanded}),
we obtain
\begin{align}
    a_J=\frac{M_{\rm r}g_J^2}{6\pi}\left(E_{\Phi,J}-\frac{M_{\rm r}g_J^2\Lambda_J^3}{12\pi}\right)^{-1},
\end{align}
\begin{align}
    r_J=-\frac{6\pi}{M_{\rm r}^2g_J^2}+\frac{24\pi E_{\Phi,J}}{M_{\rm r}g^2\Lambda_J^2}-3\Lambda_J.
\end{align}
In this study, we focus on the empirically known $J^{\pi}=3/2^{-}$ resonance with $a_{3/2}=-67.1$ fm$^3$ and $r_{3/2}=-0.87$ fm$^{-1}$~\cite{PhysRevC.102.055802}.
Here, we set $\Lambda_{3/2}=0.9$ fm$^{-1}$ 
in such a way that
the resonance energy $E_{\rm res.}$ obtained from ${\rm Re}[f_{\ell=1,J=3/2}(k=\sqrt{2M_{\rm r}E_{\rm res.}})]=0$ is close to $0.94$ MeV~\cite{10.1143/PTP.61.1327}.
In this way, we can determine $E_{\Phi,3/2}=449.607$ MeV and $M_{\rm r}g_{3/2}^2=113.388$ fm$^2$ from the empirical values of
$a_{3/2}$ and $r_{3/2}$.
These parameter sets lead to $E_{\rm res.}\simeq 0.93$ MeV.
We note that the model parameters for $J=1/2$ can also be determined in such a way that the corresponding neutron-alpha phase shift is reproduced~\cite{10.1143/PTP.61.1327,PhysRevLett.99.022502}. 

\section{In-medium $^5$He state}
\label{sec:4}
\begin{figure}[t]
    \centering
    \includegraphics[width= 0.6\hsize]{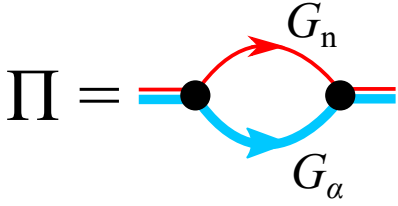}
    \caption{The diagrammatic representation of the self-energy. The thin and thick lines denoted by $G_{\rm n}$ and $G_{\alpha}$ represent the Green's functions of a neutron and an alpha particle, respectively.
    The black dots show the Feshbach-like neutron-alpha-$^5$He coupling $\sqrt{\frac{4\pi}{3}}qg_{q,J}Y_{m}^{\ell=1}
    \langle 1,m;1/2,s_z|J,J_z\rangle
    $.}
    \label{fig:2}
\end{figure}
In this section, we explain how the medium effect is incorporated in the 
description of the $^5$He state in neutron matter.
The thermal dressed propagator of $^5$He is given by
\begin{align}
    D_{J,J_z}^{\rm m}(\bm{P},i\Omega_n)&=\frac{1}{[D_{0,J}(\bm{P},i\Omega_n)]^{-1}-\Pi_{J,J_z}^{\rm m}(\bm{P},i\Omega_n)},
\end{align}
where $\Omega_n=(2n+1)\pi T$ is the fermionic Matsubara frequency.
$D_{0,J}(\bm{P},i\Omega_n)=(i\Omega_n-\xi_{\bm{P},\Phi,J}-E_{\Phi,J})^{-1}$
is the bare propagator.
The medium effect is described by the self-energy $\Pi_{J,J_z}^{\rm m}(\bm{P},i\Omega_n)$, which is diagrammatically given by Fig.~\ref{fig:2}.
Their explicit forms read
\begin{widetext}
\begin{align}
\Pi_{3/2,3/2}^{\rm m}(\bm{P},i\Omega_n)
&=-\frac{4\pi}{3}T\sum_{\bm{q},i\omega_{n'}}
q^2g_{q,3/2}^2
|Y_{1}^1(\hat{\bm{q}})|^2
G_{\rm n}(\bm{P}/2+\bm{q},i\Omega_n+i\omega_{n'})
G_{\alpha}(\bm{P}/2-\bm{q},-i\omega_{n'}),\\
%\end{align}
%\begin{align}
\Pi_{3/2,1/2}^{\rm m}(\bm{P},i\Omega_n)
&=-\frac{4\pi}{3}T\sum_{\bm{q},i\omega_{n'}}
q^2g_{q,3/2}^2
\left(\frac{1}{3}|Y_{1}^1(\hat{\bm{q}})|^2
+\frac{2}{3}|Y_{0}^1(\hat{\bm{q}})|^2
\right)
G_{\rm n}(\bm{P}/2+\bm{q},i\Omega_n+i\omega_{n'})
G_{\alpha}(\bm{P}/2-\bm{q},-i\omega_{n'}),\\
%\end{align}
%\begin{align}
\Pi_{3/2,-1/2}^{\rm m}(\bm{P},i\Omega_n)
&=-\frac{4\pi}{3}T\sum_{\bm{q},i\omega_{n'}}
q^2g_{q,3/2}^2
\left(\frac{2}{3}|Y_{0}^1(\hat{\bm{q}})|^2
+\frac{1}{3}|Y_{-1}^{1}(\hat{\bm{q}})|^2
\right)
G_{\rm n}(\bm{P}/2+\bm{q},i\Omega_n+i\omega_{n'})
G_{\alpha}(\bm{P}/2-\bm{q},-i\omega_{n'}),\\
%\end{align}
%\begin{align}
\Pi_{3/2,-3/2}^{\rm m}(\bm{P},i\Omega_n)
&=-\frac{4\pi}{3}T\sum_{\bm{q},i\omega_{n'}}
q^2g_{q,3/2}^2
|Y_{-1}^1(\hat{\bm{q}})|^2
G_{\rm n}(\bm{P}/2+\bm{q},i\Omega_n+i\omega_{n'})
G_{\alpha}(\bm{P}/2-\bm{q},-i\omega_{n'}),\\
%\end{align}
%\begin{align}
\Pi_{1/2,1/2}^{\rm m}(\bm{P},i\Omega_n)
&=-\frac{4\pi}{3}T\sum_{\bm{q},i\omega_{n'}}
q^2g_{q,1/2}^2
\left(\frac{1}{3}|Y_{0}^1(\hat{\bm{q}})|^2+\frac{2}{3}
|Y_{1}^1(\hat{\bm{q}})|^2
\right)
G_{\rm n}(\bm{P}/2+\bm{q},i\Omega_n+i\omega_{n'})
G_{\alpha}(\bm{P}/2-\bm{q},-i\omega_{n'}),\\
%\end{align}
%\begin{align}
\Pi_{1/2,-1/2}^{\rm m}(\bm{P},i\Omega_n)
&=-\frac{4\pi}{3}T\sum_{\bm{q},i\omega_{n'}}
q^2g_{q,1/2}^2
\left(\frac{1}{3}|Y_{0}^1(\hat{\bm{q}})|^2+\frac{2}{3}
|Y_{-1}^1(\hat{\bm{q}})|^2
\right)
G_{\rm n}(\bm{P}/2+\bm{q},i\Omega_n+i\omega_{n'})
G_{\alpha}(\bm{P}/2-\bm{q},-i\omega_{n'}),
\end{align}
\end{widetext}
where $G_{\rm n}(\bm{p},i\Omega_n)$ and $G_\alpha(\bm{p},i\omega_{n'})$ are the thermal Green's functions of a neutron and an alpha particle, respectively ($\omega_{n'}=2n'\pi T$ is the boson Matsubara frequency).
In this study, we consider non-superfluid neutrons for simplicity, as given by $G_{\rm n}(\bm{p},i\Omega_n)=(i\Omega_n-\xi_{\bm{p},\nu})^{-1}$.
Note that for superfluid neutrons, we can use the Nambu-Gor'kov Green's function $G_{\rm n}(\bm{p},i\Omega_n)=u_{\bm{p}}^2/(i\Omega_n-E_{\bm{p}})+v_{\bm{p}}^2/(i\Omega_n+E_{\bm{p}})$ with $E_{\bm{p}}=\sqrt{\xi_{\bm{p},\nu}^2+\Delta_{\bm{p}}^2}$,
where $\Delta_{\bm{p}}$ is the neutron pairing gap, and
$u_{\bm{p}}^2=\frac{1}{2}\left(1-\frac{\xi_{\bm{p},\nu}}{E_{\bm{p}}}\right)$ and 
$v_{\bm{p}}^2=\frac{1}{2}\left(1+\frac{\xi_{\bm{p},\nu}}{E_{\bm{p}}}\right)$
are the coherence factors~\cite{OHASHI2020103739}.
Although alpha particles may also form a condensate,
in this paper, we consider the single $^5$He state and hence the single alpha particle described by $G_\alpha(\bm{p},i\omega_{n'})=(i\omega_{n'}-\xi_{\bm{p},\alpha})^{-1}$ with $\mu_\alpha=0$.
In such a case, the summation of the Matsubara frequency can be performed analytically.
For example, we obtain
\begin{align}
    \Pi_{3/2,3/2}^{\rm m}(\bm{P},i\Omega_n)
    &=\sum_{\bm{q}}\frac{q_{z}^2g_{q,J}^2[1-f(\xi_{\bm{P}/2+\bm{q},\nu})]}{i\Omega_n-\xi_{\bm{P}/2+\bm{q},\nu}-\xi_{\bm{P}/2-\bm{q},\alpha}},
\end{align}
where the Fermi distribution function $f(\xi)=1/(e^{\xi/T}+1)$ is replaced by the step function $\theta(-\xi)$ at $T=0$.

We note that, in general, the self-energy may be in the tensor form with the off-diagonal component associated with the coupling between different $J_z$.
However, such a component does not appear due to the orthogonality of $Y_{m}^{\ell}(\hat{\bm{q}})$.
More explicitly, for the momentum integration of $\bm{q}$, we take $z$ axis along the $\bm{P}$ direction as $\bm{P}=(0,0,P)$ and $\bm{q}=(q\sin\theta\cos\phi,q\sin\theta\sin\phi,q\cos\theta)$, without loss of generality.
In this case, we obtain $\bm{q}\cdot\bm{K}=qK\cos\theta$, which is independent of $\phi$.
Therefore, we find
\begin{align}
\sum_{\bm{q}}q_iq_jF(P,q,\theta)=\frac{\delta_{ij}}{(2\pi)^2}\int_0^{\infty}dq\int_0^{\pi}d\theta q^2\sin\theta F(P,q,\theta)
\end{align}
for an arbitrary function $F(P,q,\theta)$.
Note that $[Y_{m}^{\ell=1}(\hat{\bm{q}})]^*Y_{m'}^{\ell=1}(\hat{\bm{q}})$ ($m\neq m'$) is associated with $q_iq_j$ ($i\neq j$).
In this regard, the off-diagonal component of the self-energy  disappears after the $\phi$ integration.
This fact is consistent with Ref.~\cite{PhysRevA.85.053628}.
%We can rewrite the self-energy tensor as $\Pi_{ij}^{\rm m}(\bm{P},i\Omega_n)=\delta_{ij}\Pi_{\rm d}^{\rm m}(\bm{P},i\Omega_n)$.

%\section{In-medium $^5{\rm He}$ resonance}
%\label{sec:3}
Using the in-medium propagators,
we shall examine excitation properties of the in-medium $^5$He state.
The excitation spectrum of the in-medium $^5$He state can be obtained from $D_{J,J_z}^{\rm m}(\bm{P},i\Omega_n\rightarrow \Omega_{+})$ with the analytic continuation.
We define the spectral function as
\begin{align}
    A_{J,J_z}(\bm{P},\Omega)=-\frac{1}{\pi}{\rm Im}D_{J,J_z}^{\rm m}(\bm{P},i\Omega_n\rightarrow \Omega_+).
\end{align}

In what follows, we consider the zero center-of-mass state $\bm{P}=\bm{0}$ at $T=0$.  Also, we take $\mu_\nu=E_{\rm F}$, where $E_{\rm F}=\frac{k_{\rm F}^2}{2M_\nu}$ and $k_{\rm F}$ are the Fermi energy and momentum of neutrons, respectively.
While $\Pi_{J,J_z}(\bm{P},i\Omega_n)$ depends on $J_z$ for nonzero $\bm{P}$,
we can impose the spherical symmetry of $\bm{q}$ at $\bm{P}=\bm{0}$ as in the in-vacuum case.
In this way, we obtain the $J_z$-independent self-energy at $\bm{P}=\bm{0}$ as $\Pi_{0,J}^{\rm m}(i\Omega_n)\equiv \Pi_{J,J_z}(\bm{P}=\bm{0},i\Omega_n)$.
Using this fact, we obtain
\begin{align}
    \Pi_{J}^{\rm m}(i\Omega_n\rightarrow\Omega_+)
    &=\frac{1}{3}\sum_{\bm{q}}\frac{q^2g_{q,J}^2[1-f(\xi_{\bm{q},\nu})]}{\Omega_+-\xi_{\bm{q},\nu}-\xi_{-\bm{q},\alpha}},
\end{align}
which reproduces the in-vacuum result given by Eq.~(\ref{eq:piv}) when $f(\xi_{\bm{q},\nu})\rightarrow 0$ and  $\mu_\nu\rightarrow 0$. 
One can perform the momentum integration of $\Pi_{J}^{\rm m}(\Omega_+)$ analytically as
\begin{widetext}
\begin{align}
\label{eq:pim}
     \Pi_{J}^{\rm m}(\Omega_+)
    %&=\frac{g^2}{3}\sum_{\bm{q}}
    %\gamma_q^2q^2
    %\frac{\theta(-\xi_{\bm{q},\nu})}{\Omega_+-\xi_{\bm{q},\nu}-\xi_{\bm{q},\alpha}}\cr
    &=-\frac{M_{\rm r}g_J^2\Lambda_J^4}{6\pi^2}
    \Biggl[\frac{i[2M_{\rm r}(\Omega_++E_{\rm F})]^{3/2}}{[2M_{\rm r}(\Omega_++E_{\rm F})+\Lambda_J^2]^2}
    \left\{
    %\tan^{-1}\left(\frac{k}{\sqrt{-2M_{\rm red.}(\Omega_++E_{\rm F})}}\right)
    \pi+i\ln\left(\frac{\sqrt{2M_{\rm r}(\Omega_++E_{\rm F})}-k_{\rm F}}{\sqrt{2M_{\rm r}(\Omega_++E_{\rm F})}+k_{\rm F}}\right)
    \right\}\Biggr.\cr
    &\Biggl.
    +
    \frac{k_{\rm F}\Lambda_J^2}{[2M_{\rm r}(\Omega_++E_{\rm F})+\Lambda_J^2](k_{\rm F}^2+\Lambda_J^2)}
    +
    \frac{6M_{\rm r}(\Omega_++E_{\rm F})\Lambda_J+\Lambda_J^3}{[2M_{\rm r}(\Omega_++E_{\rm F})+\Lambda_J^2]^2}
    \left\{\frac{\pi}{2}-
    \tan^{-1}\left(\frac{k_{\rm F}}{\Lambda_J}\right)
    \right\}\Biggr].
\end{align}
\end{widetext}
We note that Eq.~(\ref{eq:pim}) reproduces Eq.~(\ref{eq:piv}) in the dilute limit ($k_{\rm F}\rightarrow 0$).
This is because the diagram shown in Fig.~\ref{fig:2} is 
equivalent to the
in-vacuum case. %where there is no backward (hole-like) propagation.
In this way, we can get 
\begin{align}
    D_{J,J_z}^{\rm m}(\bm{0},\Omega_+)=\frac{1}{\Omega_+-E_{\Phi,J}+\mu_\Phi-\Pi_{J}^{\rm m}(\Omega_+)},
\end{align}
where $\mu_\Phi\equiv \mu_\nu+\mu_\alpha=E_{\rm F}$.
For convenience, we introduce
the $J_z$-independent spectral function at $\bm{P}=\bm{0}$ as $A_{J}(\Omega)\equiv A_{J,J_z}(\bm{P}=\bm{0},\Omega)$. 

\begin{figure}[t]
    \centering
    \includegraphics[width= 0.95\hsize]{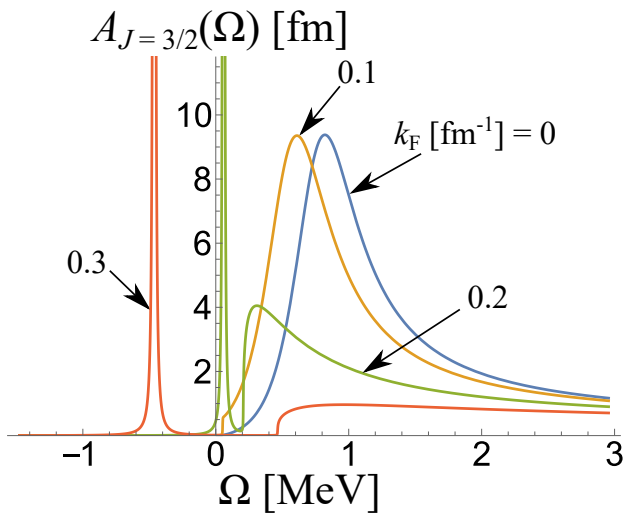}
    \caption{Spin-$3/2^{-}$ $^5$He spectral function $A_{J=3/2}(\Omega)$ in dilute neutron matter at different neutron densities (where the Fermi momentum is given by $k_{\rm F}=0$ fm$^{-1}$, $0.1$ fm$^{-1}$, $0.2$ fm$^{-1}$, and $0.3$ fm$^{-1}$ in each curve).}
    \label{fig:3}
\end{figure}

Figure~\ref{fig:3} shows the calculated spectral function $A_{J=3/2}(\Omega)$ of the $J^{\pi}=3/2^{-}$ $^5$He state at the center-of-mass frame in dilute neutron matter.
For visibility of the spectral function, we have taken $\delta=10^{-3}$~MeV.
The result at $k_{\rm F}=0$ corresponds to the $^5$He resonance in vacuum.
Indeed, the resonance peak can be found around $\Omega=E_{\rm res.}$.
As $k_{\rm F}$ increases, the resonance peak moves to the lower energy side and eventually the continuum edge appears at $\Omega=\frac{M_\nu}{M_\alpha}E_{\rm F}$ (note that it corresponds to $\Omega+E_{\rm F}=\frac{k_{\rm F}^2}{2M_{\nu}}+\frac{k_{\rm F}^2}{2M_{\alpha}}$ with the zero center of-mass momentum $\bm{P}=-\bm{k}_{\rm F}+\bm{k}_{\rm F}=\bm{0}$).
In this regard, the resonance remains broad when the peak is located above the continuum edge.
When the peak is lowered below the continuum edge, the resonance turns into the bound state exhibiting a sharp peak structure~\footnote{Note that this condition would be met only when an alpha particle happens to move with the momentum opposite to that of the partner neutron (i.e., zero center-of-mass configuration of the $^5$He state).  For an alpha particle at rest, a little bit higher density of neutrons is required for such a resonance-to-bound transition}.

\begin{figure}[t]
    \centering
    \includegraphics[width= 0.95\hsize]{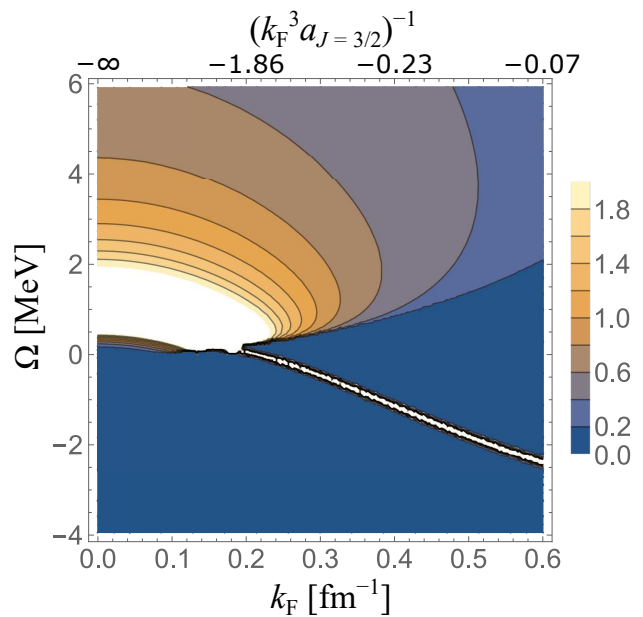}
    \caption{Contour plot of the $J^{\pi}=3/2^{-}$ $^5$He spectral function $A_{J=3/2}(\Omega)$ in the unit of fm in the plane of the energy $\Omega$ and the neutron Fermi momentum $k_{\rm F}$. On the top of the figure, we show the value of the dimensionless coupling parameter $(k_{\rm F}^3a_{J=3/2})^{-1}$ in terms of the $p$-wave scattering volume, which is frequently used in cold atomic physics.}
    \label{fig:4}
\end{figure}

In Fig.~\ref{fig:4},
we show the contour plot of $A_{J=3/2}(\Omega)$ in the plane of $\Omega$ and $k_{\rm F}$.
As shown in Fig.~\ref{fig:3}, the resonance peak appears around $\Omega=E_{\rm res.}$.
While the continuum edge moves towards the higher energy as $\frac{M_\nu}{M_\alpha}E_{\rm F}\propto k_{\rm F}^2$, 
the broad $^5$He peak is lowered gradually and changes into the bound-state peak around $k_{\rm F}=0.2$ fm$^{-1}$.
This change, which is reminiscent of the resonance-to-bound transition of the $^5$He state, occurs at a density close to the point
where $E_{\rm F}$ exceeds $E_{\rm res.}$~\cite{Note1}. %~\footnote{Note that this condition would be met only when an alpha particle happens to move with the momentum opposite to that of the partner neutron (i.e., zero center-of-mass configuration of the $^5$He state).  For an alpha particle at rest, a little bit higher density of neutrons is required for such a resonance-to-bound transition}. 
The corresponding density $\rho\simeq 2.7\times 10^{-4}$ fm$^{-3}$ for the transition
is sufficiently low compared to
the threshold $\rho\simeq 0.01$ fm$^{-3}$ of the alpha-particle condensation in asymmetric nuclear matter~\cite{PhysRevC.82.034322}.
In other words, while we 
ignore the Pauli constraint on
neutrons inside an alpha particle
by considering a point-like alpha particle, 
our description may be qualitatively valid up to $\rho\simeq 0.01$ fm$^{-3}$ (i.e., $k_{\rm F}\simeq 0.67$ fm$^{-1}$).
This value is also close to the density where $E_{\rm F}$ reaches the alpha binding energy per nucleon $\sim 7$ MeV~\cite{PhysRevC.102.055802}.
Even above this density, there is a possibility that alpha-particle-like states may remain as Cooper quartets~\cite{kamei2005quartet,BARAN2020135462,PhysRevC.105.024317}.

We conclude this section by discussing the relation between in-medium $^5$He bound states and heteronuclear Feshbach molecules.
In cold atomic systems near a narrow $p$-wave Feshbach resonance,
one can tune the $p$-wave scattering volume $a$ 
by applying an external magnetic field~\cite{RevModPhys.82.1225}.
If $a$ is tuned to be large at fixed density (i.e., $k_{\rm F}$), the $p$-wave interaction is enhanced and hence the molecular binding energy becomes large as reported in Ref.~\cite{PhysRevLett.109.075302}.
In this case, the interaction strength can be measured by the dimensionless coupling parameter $(k_{\rm F}^3a)^{-1}$.
To see the same physics in the present nuclear system,
we also show $(k_{\rm F}^3a_{J=3/2})^{-1}$ as another horizontal axis of Fig.~\ref{fig:4}.
While $a_{J=3/2}$ is unchanged in this system,
$(k_{\rm F}^3a_{J=3/2})^{-1}$ approaches zero from the negatively large value with increasing $k_{\rm F}$. At the same time, 
the dimensionless range parameter $r_{J=3/2}/k_{\rm F}$ also runs from $-\infty$ to $-1.45$. 
Therefore, one can understand the decrease of the $^5$He bound state energy by analogy with the case of the Feshbach resonance.
On the contrary, one may expect that the resonance-to-bound transition can also be found in ultracold atomic systems~(e.g., a Bose-Fermi mixture recently observed in Refs.~\cite{PhysRevA.104.043321,duda2022long}) by measuring the spectral function of the in-medium Feshbach molecule in the so-called weak-coupling Bardeen-Cooper-Schrieffer regime ($a<0$).
Indeed, the spectral function of the Feshbach molecules can be measured by the radio-frequency spectroscopy (e.g., Refs.~\cite{PhysRevLett.109.085301,PhysRevA.104.043321}).
%Note that, for the cold atomic case
%it is useful to normalize the molecular energy by using the Fermi energy as $\Omega/E_{\rm F}$, while we use the unit of ${\rm MeV}$ for the energy scale in the present nuclear system.

\section{Summary}
\label{sec:5}
In this paper, we theoretically examined the resonance-to-bound transition of the $J^{\pi}=3/2^{-}$ $^5$He ground state in dilute neutron matter. 
Using the two-channel model 
developed for
the description of a cold atomic gas near the narrow Feshbach resonance,
we succeeded in describing the $^5$He resonant state in vacuum.
Within this framework, we calculated the $^5$He spectral function in dilute neutron matter at $T=0$.
As a result of increase in the neutron density, the resonance-to-bound transition of the $^5$He state was found to occur around $k_{\rm F}=0.2$ fm$^{-1}$.
At higher densities, we have found that the spectral function exhibits a sharp peak at negative energy because the decay process to a neutron and an alpha particle is forbidden by the Pauli-blocking effect of neutrons.

For future work, we need to consider finite temperature effects and superfluid fermions for a more realistic description of astrophysical situations.
A $^5$He state with nonzero center-of-mass momentum will have to be taken into acount to discuss a possible transition from the $p$-wave alpha polaronic to $^5$He bound state in neutron matter and also the fate of multi-alpha cluster states such as $^9$Be.
For better description of such states, it would be desirable to simultaneously consider the polaronic properties of alpha particles due to the neutron-alpha $s$-wave interaction, in addition to the $p$-wave resonant or bound state of $^5$He.
The coupling with the $d$-$^3$He state would also be important for the description of the fusion reaction with spin-$3/2^+$~\cite{PhysRevC.89.014622}.

\begin{acknowledgements}
The authors thank J. Takahashi and K. Ochi for useful discussions.
H. T. thanks H. Liang and T. Naito for giving useful comments.
This work is supported by the JSPS Grants-in-Aid for Scientific Research under Grant Nos.~18H05406,~18K03635,~22K13981.
We acknowledge the Collaborative Research
Program 2022, Information Initiative Center, Hokkaido University.
\end{acknowledgements}

\bibliographystyle{apsrev4-2}
\bibliography{reference.bib}

\end{document}